\documentclass[pra,10pt,a4paper,twocolumn,showkeys,notitlepage]{revtex4}

\usepackage{times}
\usepackage{amsmath}
\usepackage{amssymb}
\usepackage[dvipdf]{graphicx}

\begin{document}

\title[On a heuristic point of view related to quantum nonequilibrium
statistical mechanics]{On a heuristic point of view related to quantum
nonequilibrium statistical mechanics}
\author{Norton G. de Almeida}
\email{norton@pq.cnpq.br}
\affiliation{N\'{u}cleo de Pesquisas em F\'{\i}sica, Universidade Cat\'{o}lica de Goi\'{a}%
s, 74.605-220, Goi\^{a}nia (GO), Brazil.}
\pacs{PACS number}

\begin{abstract}
In this paper I propose a new way for counting the microstates of a system
out of equilibrium. As, according to quantum mechanics, things happen as if
a given particle can be found in more than one state at once, I extend this
concept to propose the coherent access by a particle to the available states
of a system. By coherent access I mean the possibility for the particle to
act as if it is populating more than one microstate at once. This hypothesis
has experimental implications, since the thermodynamical probability and, as
a consequence, the Bose-Einstein distribution as well as the argument of the
Boltzmann factor is modified.
\end{abstract}

\maketitle

\section{Introduction}

Nowadays, the development of new entropic forms has been followed by an
increasing interest, as can be seen, for example, in Refs.\cite%
{nonextensive,nonextensive1}. Although it is possible to formulate new
entropies from a strictly mathematical point of view \cite{nonextensive},
without connection with the physics implicated in such formulations,
recently some works have appeared trying to understand the link between the
physical situation and the mathematical formulation \cite{Wilk00,Beck01}. In
this meantime, some works have appeared rising the question of a possible
pseudononextensivity stemming from the generalized entropic forms \cite%
{Planes02,Norton08}.

In general, nonextensive formulations are related to nonequilibrium
situations, where the Boltzmann factor $\exp (-\beta E)$, presumably, plays
not a preponderant role, being difficult, if not impossible, to associate a
definite temperature to the system. In some cases, however, by considering
situations only slightly out of the equilibrium, it is possible to ascribe a
temperature to the system, which results in a distribution function
different from that of Boltzmann \cite{Clayton74}.

In this paper, inspired by some ideas from quantum optics, I propose a new
way for counting accessible states to a given particle, in such a way that
its thermodynamical probability $\Omega $\ is modified, with direct
consequences in the entropic form $S\propto \Omega $\ of the system. As is
well known in the quantum optics domain, which deals fundamentally with
nonequilibrium systems, an initially pure state can be described, in its
most general form, as a superposition of each state physically accessible to
the particle. The role of the reservoir, even at the idealized zero
temperature, is to lead the system to a complete mixture at the end of the
so-called decoherence time $\tau _{D}$. Thus, even before the thermalization
occurs the loss of coherence of the system, or, in other words, the system
capacity to access, coherently, every possible state. This state of affairs
suggests an entirely new way to count the accessible states. It is this
connection, until now not explored, between the new way to count the
accessible states and its consequences to the the entropic form of the
system that we will explore in the next sections.

\section{Redefining the microstates}

In statistical mechanics, to define a microstate it is necessary to take
into account the (un)distinguishability of the particles, which gives rise
to different configurations (see Tab.1). 
\begin{table}[h]
\begin{tabular}[b]{||c||c||}
\hline\hline
(1) & (2) \\ \hline
$a$ & $a$ \\ \hline
$aa$ &  \\ \hline
& $aa$ \\ \hline\hline
\end{tabular}
(a)\ \ \ \ \ \ \ \ \ \ \ \ \ \ 
\begin{tabular}[b]{||c||c||}
\hline\hline
(1) & (2) \\ \hline\hline
$a$ & $b$ \\ \hline
$b$ & $a$ \\ \hline
$ab$ &  \\ \hline
& $ab$ \\ \hline\hline
\end{tabular}
(b)
\caption{The configuration of two accessible states for a) two
indistinguishable particles and b) two distinguishable particles. (1)
denotes the first available state and (2) denotes the second available
state. }
\label{table1}
\end{table}
For calculating all the possible configurations we now take into account,
beside this characteristic, this another one: the possibility to the
particle simultaneously access more than one state, or, to avoid eventual
difficulties related to interpretations matter inherent to the quantum
formalism, the possibility to the particle to coherently access the
available states. This situation is shown in Tab. 2 for the case of two
identical particles having two accessible states. Note that if the particles
are distinguishable, the corresponding configuration is different.

\begin{table}[h]
\begin{tabular}[b]{||c||c||c||}
\hline\hline
(1) & (2) & (12) \\ \hline\hline
$\circ $ & $\circ $ &  \\ \hline
$\circ $ &  & $\circ $ \\ \hline
& $\circ $ & $\circ $ \\ \hline
$\circ \circ $ &  &  \\ \hline
& $\circ \circ $ &  \\ \hline
&  & $\circ \circ $ \\ \hline\hline
\end{tabular}%
\caption{A system out of equilibrium composed by two particles having two
accessible states. (1) denotes the first available state, (2) denotes the
second available state, and (12) denotes the coherent access to both states.}
\label{Tab2}
\end{table}

Comparing Tab.I and Tab.II, we see that, clearly, the nonequilibrium
situation requires a new way for counting microstates. This new way to
count, shown in Tab.2, can be represented by the following sequences, where
the number between parentheses indicates the state occupied and the letter
following the parenthesis indicates the corresponding occupation by the
particle $a$, which is identical to all the others: 
\begin{eqnarray}
&&(1)a(2)a(12);(1)a(2)(12)a;(1)(2)a(12)a;  \notag \\
&&(1)aa(2)(12);(1)(2)aa(12);(1)(2)(12)aa.  \label{Ctab2}
\end{eqnarray}%
As for example, the last sequence, $(1)(2)(12)aa$, corresponds to two
particles accessing coherently the two states $(1)$ and $(2)$, while the
second sequence, $(1)a(2)(12)a$ corresponds to one particle accessing the
state $(1)$ and the other accessing coherently the states $(1)$ and $(2)$.
Note that, as the sequence must initiate by a number, and existing three
possible number of states, $1,2,$ and $12$, will remain $3-1$ numbers plus
two letters $a$ (particles) to be set in whatever order (permutation).
Therefore, the number of unrepeated sequences is%
\begin{equation}
w^{\ast }=\frac{3\times (3-1+2)!}{2!3!}=6\text{,}
\end{equation}%
where we have put a superscript ($\ast $)\ to remind us that we are treating
with nonequilibrium situation. Proceeding in a general manner, for $g_{j}$
sublevels with $N_{j}^{\ast }$ particles, the number $w_{j}^{\ast }$ of
unrepeated sequences is%
\begin{equation}
w_{j}^{\ast }=\frac{G_{j}(G_{j}+N_{j}^{\ast }-1)!}{G_{j}!N_{j}^{\ast }!}=%
\frac{(G_{j}+N_{j}^{\ast }-1)!}{(G_{j}-1)!N_{j}^{\ast }!}\text{,}
\label{PBMS}
\end{equation}%
where $G_{j}=\sum_{k=1}^{g_{j}}C_{g_{j}\text{,}k}$ is the number of possible
sequences formed from $g_{j}$, and $C_{n\text{,}m}=n!/(n-m)!m!$. Taking as
example the configuration given by Tab.2, where $g_{j}=1$, $%
N_{j}=2,G_{j}=\sum_{k=1}^{2}C_{2\text{,}k}$, thus $G_{j}=C_{2\text{,}1}+C_{2%
\text{,}2}=3$; then%
\begin{equation}
w_{j}^{\ast }=\frac{(3+2-1)!}{(3-1)!2!}=6\text{,}
\end{equation}%
which is the number of sequences given in Eq.(\ref{Ctab2}) corresponding to
Tab.2. Therefore, the nonequilibrium thermodynamical probability $%
w_{k}^{\ast }$ for a given macrostate $k$ is%
\begin{equation}
w_{k}^{\ast }=\prod\limits_{j}\frac{(G_{j}+N_{j}^{\ast }-1)!}{%
(G_{j}-1)!N_{j}^{\ast }!}.  \label{PBT}
\end{equation}%
As $G_{j}=\sum_{k=1}^{g_{j}}C_{g_{j}\text{,}k}=C_{g_{j}\text{,}%
1}+\sum_{k=2}^{g_{j}}C_{g_{j}\text{,}k}$, and $C_{g_{j}\text{,}1}=g_{j}$,
letting $L_{gj}=\sum_{k=2}^{g_{j}}C_{g_{j}\text{,}k},$ then Eq.(\ref{PBT})
can be written as%
\begin{equation}
w_{k}^{\ast }=\prod\limits_{j}\frac{(g_{j}+L_{gj}+N_{j}^{\ast }-1)!}{%
(g_{j}+L_{gj}-1)!N_{j}^{\ast }!}\text{.}  \label{PBT1}
\end{equation}%
From Eq.(\ref{PBT1}) we can see that the only changing in the
thermodynamical probability is the appearance of the factor $L_{gj}$
modifying the degeneracy $g_{j}$, and, as a consequence, modifying also the
number of macrostates, $\Omega =\sum_{k}w_{k}^{\ast }$, and the entropy of
the system. Before ending this section, we call attention to the
plausibility in presume that, given a system out of the equilibrium with $%
N^{\ast }$ particles and $n$ levels, each of this having $g_{j}$ sublevels,
as the equilibrium is established ($L_{gj}\rightarrow 0$), the $N^{\ast }$
particles of the system accommodate by the $n_{j}$ levels, with each level
receiving $N_{j}^{\ast }$ particles, which are distributed by the sublevels.
Also, as it is easily verified, Eq.(\ref{PBT1}) gives rise to a
Bose-Einstein-like statistics, with $g_{j}$ replaced by $G_{j}$. That this
is so can be checked in the following manner, proceeding by analogy with the
equilibrium situation: First, we take the $\ln $ from both sides of Eq.(\ref%
{PBMS}). Second, we use the Stirling formula. Third, we differentiate with
respect to $N_{j}^{\ast }$ and use $\partial \ln w_{j}^{\ast }/\partial
N_{j}^{\ast }=\epsilon _{j}^{\ast }$ , where $\epsilon _{j}^{\ast }$
generalizes $\varepsilon _{j}=\beta E_{j}$, $\beta $ $=1/kT$ ,$\ $to find 
\begin{equation}
N_{j}^{\ast }/G_{j}=\frac{1}{\exp (\epsilon _{j}^{\ast })-1}.  \label{BElike}
\end{equation}%
This is an interesting point, having experimental implication: the
Bose-Einstein statistics is corrected, since the equality $G_{j}=g_{j}$ and $%
\epsilon _{j}^{\ast }=\varepsilon _{j}$ will be valid only when the complete
equilibrium is reestablished. Thus, for systems only slightly out of the
equilibrium, the energy emitted should be slightly different from that
corresponding to the system in equilibrium. Note that, as $\epsilon
_{j}^{\ast }=\varepsilon _{j}=\beta E_{j}$ when the equilibrium is restated,
it is convenient to expand $\epsilon _{j}^{\ast }$ in power series of $%
\varepsilon $%
\begin{equation}
\epsilon _{j}^{\ast }=\epsilon _{0}^{\ast }+\frac{\partial \epsilon
_{j}^{\ast }}{\partial \varepsilon _{j}}\varepsilon _{j}+\frac{1}{2!}\frac{%
\partial ^{2}\epsilon _{j}^{\ast }}{\partial \varepsilon _{j}^{2}}%
\varepsilon _{j}^{2}+\frac{1}{3!}\frac{\partial ^{3}\epsilon _{j}^{\ast }}{%
\partial \varepsilon _{j}^{3}}\varepsilon _{j}^{3}...\text{,}  \label{exp}
\end{equation}%
which, requiring that $\epsilon _{j}^{\ast }\rightarrow \varepsilon
_{j}=\beta E_{j}$ when the equilibrium is restated, gives $\epsilon
_{0}^{\ast }=0$ and $\frac{\partial \epsilon _{j}^{\ast }}{\partial
\varepsilon _{j}}=1$, such that the first order correction to the
Bose-Einstein distribution can be explicitly written as%
\begin{equation*}
N_{j}^{\ast }/G_{j}=\frac{1}{\exp \left[ \beta E+\alpha _{1}\left( \beta
E\right) ^{2}\right] -1}\text{,}
\end{equation*}%
where we have kept only a few terms and put $\frac{1}{2!}\frac{\partial
^{2}\epsilon _{j}^{\ast }}{\partial \varepsilon _{j}^{2}}=\alpha _{1}$. Note
that from this approach the net effect stemming from the nonequilibrium on a
given system is the increasing in the degeneracy, which in turn increases
the available states given by $\Omega .$ The Boltzmann factor, to be
recovered when $\exp (\beta ^{\ast }\epsilon _{j}^{\ast })\gg 1$ , is
modified, and we will explore more about this in the next Section. The
choice of the more convenient entropic form associated to this new
thermodynamical probability is discussed in the last Section.

\section{ The nonequilibrium partition function}

Let us focus our attention to the bosonic particles, since the other cases
are similar. By definition, the partition function is defined as a sum in
all microstates ($ms$):%
\begin{equation}
Z=\sum_{ms}\exp (-\beta E)\text{,}  \label{PF}
\end{equation}%
where $E$ is the energy of the system and $\beta $ is related to the
temperature $T$ of the system by the Boltzmann constant $\beta $ $=1/kT$.
Writing the energy $E$ in terms of the number of particles \bigskip $n_{i}$
in the state of $i-$th energy $\epsilon (i)$ of the system, we will have $%
E=\sum_{i}n_{i}\epsilon (i)$. Of course, in this case the total number of
particles is simply $N=\sum_{i}n_{i}$.

For an out of equilibrium system, we introduce the coherent access
hypothesis to several states, which consists in maintaining the same form as
that of Eq.(\ref{PF}), but replacing $\sum_{i}n_{i}\epsilon (i)$ by $%
\sum_{ij...}n_{ij}\epsilon (i,j,...)$, \ where $n_{ij}$ must be interpreted
as being the number of particles coherently accessing the energy levels $%
\epsilon (i)$ and $\epsilon (j)$. For example, as discussed in Section I and
represented in Tab.2, $\epsilon (i,j)$ represents the coherent access
related to the energy levels $i$ and $j$, and $\epsilon (1,2)$ represents,
for example, the states $(1)$ and $(2)$ being coherently populated.

For demonstrating that the partition function preserves its form given by
Eq.(\ref{PF}) even at the nonequilibrium situation, it is enough to maintain
this following postulate, which is valid for equilibrium situation: that two
systems, in contact with a third one, as for example a reservoir at
temperature $T$, act independently of each other while both the systems
exchange energy with the reservoir. Although this demonstration is
straightforward, for completeness we address the reader to the appendix.
Continuing to denote the nonequilibrium quantities with a superscript ($\ast 
$), thus according to Eq.(\ref{PFNE}) of the appendix, if $P(\epsilon
_{j}^{\ast }=\beta ^{\ast }E_{j}^{\ast })$ is the probability for a given
system out of the equilibrium is in a particular microstate whose
configuration is described by $\epsilon _{j}^{\ast }=\beta ^{\ast
}E_{j}^{\ast }$, then 
\begin{equation}
P(\epsilon _{j}^{\ast })=\frac{\exp (-\epsilon _{j}^{\ast })}{Z^{\ast }}%
\text{.}  \label{NE}
\end{equation}%
Now, using Eq.(\ref{exp}) and requiring that $\epsilon _{j}^{\ast
}\rightarrow \beta E_{j}$ when the equilibrium is restated, the Eq(\ref{NE})
can now be written as 
\begin{eqnarray}
P(\epsilon _{j}^{\ast }) &=&\frac{1}{Z^{\ast }}\exp \left[ -\beta
E_{j}-\alpha _{1}\left( \beta E_{j}\right) ^{2}-\right.  \notag \\
&&\left. \alpha _{2}\left( \beta E_{j}\right) ^{3}+\alpha _{3}\left( \beta
E_{j}\right) \right] ^{4}...\text{,}  \label{BFexpand1}
\end{eqnarray}%
where the other constants were renamed for convenience as $\frac{1}{n!}\frac{%
\partial ^{n}\epsilon _{j}^{\ast }}{\partial \varepsilon _{j}^{n}}=\alpha
_{n-1}$. Such a state of affairs giving origin to an infinite number of free
parameters was studied in Refs. \cite{Tsallis1,Norton08} in a different
context. Note that for systems only slightly out of the equilibrium this
last equation can be written as%
\begin{equation}
P(E)=\frac{1}{Z}\exp \left[ -\beta E-\alpha _{1}\left( \beta E\right) ^{2}%
\right] \text{,}
\end{equation}%
where we have dropped out the superscript and the index $i$. Some
experiments seem to point for the importance of this last term, which
modifies the Boltzmann factor \cite{Clayton74}.

\section{Connection with entropic forms}

As discussed in Section I, since the thermodynamical probability was
modified, a natural question emerging is what is the best entropic form
related to it. Of course, depending on our choice we will face with
different implications. Once there is a plenty of entropic forms at our
disposal, we will focus our attention only in two of them: the
Boltzmann-Gibbs ($S_{BG}$) and the Tsallis ($S_{q}$) entropies. As is well
known, while the first is extensive, \textit{i.e}. $%
S_{BG}(A+B)=S_{BG}(A)+S_{BG}(B)$, the second in general is not, \textit{i.e}%
., $S_{q}(A+B)\neq S_{q}(A)+S_{q}(B)$ if $q\neq 1$.

Let us begin adopting the Boltzmann-Gibbs entropy, assuming for now that the
single effect of the nonequilibrium is to increase the degeneracy of the
system, as seen in Section I. It will be possible to reconcile Eq.(\ref%
{BFexpand1}) to an extensive entropic form such as that of Boltzmann and
Gibbs?\ Indeed, that this is possible was shown in Ref.\cite{Norton08}, in
the following way. \bigskip Given the density operator $\rho $ of the system
and the Boltzmann constant $k$, for maximizing the Boltzmann-Gibbs entropy $%
S_{BG}=-kTr\rho \ln \rho $ subjected to the constraints given by the moments 
\begin{equation}
\left\langle \left( \Delta E\right) ^{n}\right\rangle =Tr\rho H^{n}\text{,}
\label{2bb}
\end{equation}%
$n$ integer, we vary $\rho $ in $S_{BG}$ and in those for the constraints,
Eq.(\ref{2bb}), multiplying each constraint by the undetermined Lagrange
multiplier $\beta _{n}$, and adding the result, obtaining 
\begin{equation}
Tr\left( 1+\sum\limits_{n=0}^{\infty }\beta _{n}H^{n}+\ln \rho \right)
\delta \rho =0\text{.}
\end{equation}%
Since all the variations are independent and $\delta \rho $ is arbitrary, it
follows the extended (non-Maxwellian) distribution $\ln \rho
=-1-\sum\limits_{n=0}^{\infty }\beta _{n}H^{n}$, or, equivalently 
\begin{equation}
\rho =Z^{-1}\exp (-\sum\limits_{n=1}^{\infty }\beta _{n}H^{n})\text{,}
\label{2c}
\end{equation}%
where the partition function is $Z=Tr\exp (-\sum\limits_{n=1}^{\infty }\beta
_{n}H^{n})$. In the energy representation where $H\left\vert E\right\rangle
=E\left\vert E\right\rangle $, Eq.(\ref{2c}) now reads,%
\begin{eqnarray}
P(E) &=&Z^{-1}\exp (-\sum\limits_{n=1}^{\infty }\beta _{n}E^{n})=Z^{-1}\times
\notag \\
&&\exp \left( -\beta _{1}E+\beta _{2}E^{2}+\beta _{3}E^{3}+\beta
_{4}E^{4}...\right)  \label{2ccc}
\end{eqnarray}%
with $Z=\sum_{E}\exp (-\sum\limits_{n=1}^{\infty }\beta _{n}E^{n})$. The
Lagrange multipliers $\beta _{k}$ are formally obtained from $\beta _{k}=-%
\frac{\partial \ln Z}{\partial E^{k}}$, considering $E^{k}$ $=Y_{k}$ as
independent variables. The equality between Eq.(\ref{2ccc}) and Eq.(\ref%
{BFexpand1}) is guaranteed, provided that $\beta _{n}=\alpha _{n-1}\beta
^{n} $ and $\beta _{1}=\alpha _{0}=\beta $. Therefore, according to this
view nonequilibrium systems remains extensive, although requiring a \textit{%
posteriori} knowledge of the variance (second central moment), the
coefficient of skewness (third central moment), the kurtosis (fourth central
moment), and so on, thus giving rise virtually to an infinite number of free
parameters.

Of course, instead of using infinite parameters, we could just use a single
one by redefining \ a new ensemble fully determined by this single
parameter. An aesthetically appealing way to do so is to expand Eq.(\ref%
{BFexpand1}) in terms of the Tsallis entropic index \cite{tsallis88}, as we
will see in a moment. Consider the following expanded form of Eq.(\ref%
{BFexpand1}): 
\begin{eqnarray}
P(E_{j}) &=&\left[ \frac{1}{Z}\exp -\beta E_{j}-\frac{\left( 1-q\right) }{2}%
\left( \beta E_{j}\right) ^{2}-\right.   \notag \\
&&\left. \frac{\left( 1-q\right) ^{2}}{3}\left( \beta E_{j}\right) ^{3}+%
\frac{\left( 1-q\right) ^{3}}{4}\left( \beta E_{j}\right) ^{4}...\right] 
\text{,} \nonumber \\ \label{BFexpq}
\end{eqnarray}%
where in general $\alpha _{n}=$ $\frac{\left( q-1\right) ^{n-1}}{n}$. This
is equivalent to the statement that the old ensemble which depended of $%
\beta ,\left\{ \alpha _{n}\right\} $ and $E_{j}$\ becomes now a function of
only $\beta $, $q$ and $E_{j}$. Eq.(\ref{BFexpq}) can be rewritten as%
\begin{eqnarray}
P(E_{j}) &=&\frac{1}{Z}\exp \left\{ \frac{1}{1-q}\left[ -\left( 1-q\right)
\beta E_{j} \right. \right. \nonumber \\ &-&\frac{\left( 1-q\right) ^{2}}{2}\left( \beta E_{j}\right)^{2} - \frac{\left( 1-q\right) ^{3}}{3}\left( \beta E_{j}\right)
^{3} \nonumber \\
&-&\left. \left. \frac{\left( 1-q\right) ^{4}}{4}\left( \beta E_{j}\right) ^{4}...\right]
\right\} \text{,} \label{BFexpq1}
\end{eqnarray}%
where it is easily recognized the expanded form of the logarithm function $%
\ln (1-x)=-x-\frac{x^{2}}{2}-\frac{x^{3}}{3}-\frac{x^{4}}{4}-...$ , $%
x=(1-q)\beta E_{j}$, such that Eq.(\ref{BFexpq1}) becomes%
\begin{equation}
P(E)=\frac{1}{Z}\left[ 1-\left( 1-q\right) \beta E_{j}\right] ^{\frac{1}{%
\left( 1-q\right) }}\text{,}  \label{PTsallis}
\end{equation}%
which is the $q$-distribution stemming from the extremization of Tsallis
entropy 
\begin{equation}
S_{q}=k\frac{1-\sum\limits_{j}p_{j}^{q}}{q-1}  \label{ETsallis}
\end{equation}%
when considering a family of constraints determined by the $q$-expectation
value of the energy%
\begin{equation}
\left\langle E\right\rangle _{q}=\frac{\sum\limits_{j}p_{j}^{q}E_{j}}{%
\sum\limits_{j}p_{j}^{q}}
\end{equation}%
besides the norm constraint $\sum\limits_{j}p_{j}^{q}=1$. Therefore, a
formal agreement between Tsallis and Boltzmann-Gibbs entropies is afforded.
As pointed out in Ref.\cite{Norton08}, this formal equivalence between the
Boltzmann-Gibbs and Tsallis entropy gives rise to an important issue related
to a possible pseudononextensivity.

\section{Conclusion}

In this paper I explored an analogy between the nonequilibrium
thermodynamics and some well-established situations from quantum optics,
concerning the problem of coherent access to the multiple states available
to a given particle. As a consequence of the coherent access hypotheses, the
process of counting the possible states of a physical system is modified. I
have found a modification on both Bose-Einstein and Boltzmann-Gibbs
distribution, which is in principle experimentally detectable. Actually, it
is possible that the correction to the Boltzmann factor obtained by the
method developed here is the one suggested by some experiments \cite%
{Clayton74}. Although I have exemplified for the specific case of bosons,
the extension to fermions is straightforward. Finally, I expect that the
coherent access hypothesis introduced here eventually makes possible the
exploration of new ways of treating problems related to nonequilibrium
situations, or differing from the equilibrium in a slightly manner.

\section*{Appendix I}

To demonstrate that the partition function and the Boltzmann factor retain
the same form as Eq.(\ref{PF}) in the nonequilibrium situation, it is enough
to follow the usual derivation, as for example, that given in Ref.\cite%
{Livro}. Thus, consider a system composed by two subsystems $A$ and $B$. The
probability for this composed system to be in the energy state $%
E_{A+B}^{\ast }$ is $P_{A+B}(E_{A+B}^{\ast })$, where the superscript (*)
remind us that the system is out of equilibrium. If, as usual, the
interaction energy can be neglected, thus the energy of the composed system
is $E_{A+B}^{\ast }=E_{A}^{\ast }+E_{B}^{\ast }$, and%
\begin{equation}
P_{A+B}(E_{A+B}^{\ast })=P_{A}(E_{A}^{\ast })+P_{B}(E_{B}^{\ast })
\label{PFsum}
\end{equation}%
is the probability for the composed system to be in a particular state such
that the subsystem A has an energy $E_{A}^{\ast }$, and, at the same time,
the subsystem $B$ has an energy $E_{B}^{\ast }$. Now, suppose that these two
subsystems is put in contact with a third system, for example, a reservoir
at temperature $T$. While persisting the nonequilibrium situation (and even
after that), the two subsystems $A$ and $B$ act independently of each other,
with both subsystems eventually exchanging energy with the reservoir. Beside
that, the energy exchanged with the reservoir by a given subsystem does not
influence the energy that the other subsystem can exchange with this same
reservoir. This assumption, valid for two systems in equilibrium with a
reservoir, is here assumed to be valid also when the equilibrium was not
reached. Therefore, as these events are independent, we can write%
\begin{equation}
P(E_{A+B}^{\ast })=P(E_{A}^{\ast })P(E_{B}^{\ast })\text{.}  \label{PFdot}
\end{equation}%
Differentiating Eq.(\ref{PFdot}) with respect to $E_{A}^{\ast }$ and $%
E_{B}^{\ast }$ and equating this result we obtain ($dP/dE^{\ast }=P^{\prime
} $)%
\begin{equation}
P_{A}^{\prime }(E_{A}^{\ast })P_{B}(E_{B}^{\ast })=P_{A}(E_{A}^{\ast
})P_{B}^{\prime }(E_{B}^{\ast })\text{.}
\end{equation}%
Next, separating the variables and equating the result to a constant, we have%
\begin{equation}
\frac{P_{A}^{\prime }(E_{A}^{\ast })}{P_{A}(E_{A}^{\ast })}=\frac{%
P_{B}^{\prime }(E_{B}^{\ast })}{P_{B}(E_{B}^{\ast })}=-\beta ^{\ast }
\label{Separate}
\end{equation}%
where $\beta ^{\ast }$ is a constant independent from either $E_{A}^{\ast }$
or $E_{B}^{\ast }$. Of course, in the equilibrium situation we must have $%
\beta ^{\ast }\rightarrow \beta =1/kT$. From Eq.(\ref{Separate}) follows,
therefore, our desired result%
\begin{equation}
P(E^{\ast })=\frac{\exp (-\beta ^{\ast }E^{\ast })}{Z^{\ast }}\text{,}
\label{PFNE}
\end{equation}%
where the partition function for the nonequilibrium situation is $Z^{\ast }$
and the index were dropped given the validity of Eq.(\ref{PFNE}) for the two
subsystems.

\subsection*{Acknowledgments}

We thanks the CNPq, Brazilian Agency, and VPG - Vice-Reitoria de P\'{o}s
Gradua\c{c}\~{a}o e Pesquisa da Universidade Cat\'{o}lica de Goi\'{a}s, for
the partial support of this work.

\end{document}